\documentclass[prl,aps,twocolumn,floatfix,showpacs,tightenlines,superscriptaddress,amsmath,amssymb]{revtex4}
\usepackage{amssymb,amsbsy,epsfig,color,graphicx,times}

\newcommand{\ket}[1]{|#1\rangle}

\newcommand{\eq}[1]{Eq.~(\ref{#1})}
\newcommand{\fig}[1]{Fig.~\ref{#1}}
\newcommand{\eqs}[1]{Eqs.~(\ref{#1})}

\begin{document}

\title{Theory of ground state factorization in quantum cooperative systems}

\author{Salvatore M. Giampaolo}
\affiliation{Dipartimento di Matematica e Informatica,
Universit\`a degli Studi di Salerno, Via Ponte don Melillo,
I-84084 Fisciano (SA), Italy} \affiliation{CNR-INFM Coherentia,
Napoli, Italy; CNISM, Unit\`a di Salerno; and INFN, Sezione di
Napoli - Gruppo Collegato di Salerno, Italy}

\author{Gerardo Adesso}
\affiliation{Dipartimento di Matematica e Informatica,
Universit\`a degli Studi di Salerno, Via Ponte don Melillo,
I-84084 Fisciano (SA), Italy} \affiliation{CNR-INFM Coherentia,
Napoli, Italy; CNISM, Unit\`a di Salerno; and INFN, Sezione di
Napoli - Gruppo Collegato di Salerno, Italy}

\author{Fabrizio Illuminati}
\thanks{Corresponding author: illuminati@sa.infn.it}
\affiliation{Dipartimento di Matematica e Informatica,
Universit\`a degli Studi di Salerno, Via Ponte don Melillo,
I-84084 Fisciano (SA), Italy} \affiliation{CNR-INFM Coherentia,
Napoli, Italy; CNISM, Unit\`a di Salerno; and INFN, Sezione di
Napoli - Gruppo Collegato di Salerno, Italy} \affiliation{ISI
Foundation for Scientific Interchange, Viale Settimio Severo 65,
I-10133 Turin, Italy}


\begin{abstract}

We introduce a general analytic approach to the study of factorization
points and factorized ground states in quantum cooperative systems. The
method allows to determine rigorously existence, location, and exact form
of separable ground states in a large variety of, generally non-exactly solvable,
spin models belonging to different universality classes. The theory
applies to translationally invariant systems, irrespective of spatial
dimensionality, and for spin-spin interactions of arbitrary range.

\end{abstract}

\date{March 31, 2008}

\maketitle



{\noindent \bf Introduction}.-- Quantum engineering and quantum tech\-no\-lo\-gy
have been developing at a fast pace in recent years. Quantum devices are
being vigorously pursued for applications ranging from nano-sciences to
quantum information and entanglement-enhanced metrology \cite{QuantumEngineering}.
Despite a large variety of possible implementations involving different physical
systems, many relevant properties of such devices can be investigated in a
unified setting by appropriate mappings to quantum spin models \cite{Hartmann,Sorensen}.
Thus, control of ground state entanglement in quantum spin systems
plays an important role in quantum technology applications
\cite{Amico1}.
On the other hand, knowledge of exact solutions endowed with precisely
determined properties of separability or entanglement, can be of great
relevance in the study of advanced models of condensed matter and
cooperative systems that are in general not exactly solvable.



The occurrence of totally factorized (unentangled)
ground states of quantum many-body systems was first discovered
in the one-dimensional anisotropic Heisenberg model with
nearest-neighbor interactions \cite{Kurmann}.
This result was later re-derived and extended to two dimensions using
quantum Monte Carlo numerical methods \cite{Roscilde}.
Complex quantum systems exhibiting cooperative behaviors,
whose ground states are typically entangled \cite{Typical},
may thus admit, for some non trivial values of the
Hamiltonian parameters, a ground state which is completely separable.
The phenomenon of ground state factorization appears to be associated with the presence
of an ``entanglement phase transition'' with no classical counterpart \cite{Amico2};
furthermore, for the purposes of quantum engineering applications
that employ distributed entanglement in order to manipulate and
transfer information \cite{Bose}, factorization points need to be
exactly identified and avoided to guarantee the reliable implementation
of quantum devices. Finally, for models not admitting exact general
solutions, achieving knowledge of the exact ground state, even if only for
the restricted nontrivial set of parameters associated to factorization,
would allow {\em (i)} to prove the existence of an ordered phase
and characterize it; {\em (ii)} to build variational
or perturbative approximations around the exact factorized solution, that
may then be used as test benchmarks for the validity and the precision of
numerical algorithms and simulations.
Unfortunately, to date, it has been extremely hard to go beyond the pioneering
result of Kurmann {\em et al.} \cite{Kurmann} despite the fact that,
to prove total factorization, it would suffice to show the vanishing
of the von Neumann entropy of entanglement, or of the
linear entropy (tangle) \cite{Coffman}.
%
The difficulty resides in the fact that, with few special exceptions,
explicit analytic expressions for these measures of entanglement cannot
be obtained. Hence, the only possibility to gain insight on factorization
in systems of increasing complexity has relied so far on heuristic or
numerical approaches.

In the present work we introduce a general analytic method
that allows to determine exactly the existence of factorized ground
states and to characterize their properties in quantum spin models
defined on regular lattices, in any spatial dimension, and with
spin-spin interactions of arbitrary range. In correspondence to
rigorously established ground state factorizability, the method
also allows to determine novel sets of exact solutions in generally
non exactly solvable models. Previous particular findings for models
with short range interactions are rigorously re-derived and extended
within a unified framework inspired by concepts of quantum information
science. The method is built on a formalism
of single-spin, or single-qubit, unitary operations (SQUOs) and associated
entanglement excitation energies (EXEs), previously introduced for the
characterization and quantification of entanglement in systems of quantum
information \cite{Giampaolo1,Giampaolo2}. The novel techniques exploit the fundamental
property enjoyed by the EXEs, of vanishing if and only if a pure state is fully
factorized \cite{Giampaolo2}.
For any given Hamiltonian, the strategy to the understanding of factorization is
first to assume as working point a phase endowed with some kind of magnetic order.
Next, by imposing the vanishing of the EXE and of the linear entropy, one derives
a closed set of conditions whose solutions determine uniquely the occurrence
(or the non occurrence) of factorization points at which a quantum ground state
is completely disentangled.
Besides the rigorous determination of novel factorization points and exact
solutions of generic quantum spin models, the method allows as well to
re-derive analytically the few previously known results on ground state
factorization \cite{Kurmann,Roscilde,Dusuel}.


\vspace{0.2cm}

{\noindent \bf The method}.-- To fix ideas and notations,
let us consider general, translationally invariant, exchange
Hamiltonians for spin-$1/2$ systems on $d$-dimensional regular lattices,
with spin-spin interactions of arbitrary range and arbitrary anisotropic
couplings. This class of Hamiltonians encompasses a very large set of
models describing different spin systems and spanning several universality
classes like, among others, the Ising, XY, Heisenberg, and XYZ symmetries.
The general Hamiltonian can be written in the form

\begin{equation}\label{Hamiltonian}
 H=\frac{1}{2} \sum_{\underline{i},\underline{l}} J_x^r S_{\underline{i}}^x S_{\underline{l}}^x+
 J_y^r S_{\underline{i}}^y S_{\underline{l}}^y +J_z^r S_{\underline{i}}^z S_{\underline{l}}^z-
 h\sum_{\underline{i}} S_{\underline{i}}^z \, .
\end{equation}
Here $\underline{i}$ (and similarly $\underline{l}$) is a
$d$-dimensional index vector identifying a site in the lattice,
$S_{\underline{i}}^\alpha$ $(\alpha=x,y,z)$ stands
for the spin-$1/2$ operator on site $\underline{i}$, $h$ is external
field directed along the $z$ direction, $r=|\underline{i}-\underline{l}|$
is the distance between two lattice sites, and $J_\alpha^r$ is the spin-spin
coupling along the $\alpha$ direction; translational invariance implies that
it depends only on the distance $r$ between the spins.
Without loss of generality, one can impose
$|J_x^r|\ge |J_y^r|,|J_z^r| \; \forall r$. This condition guarantees that
at a particular value of the external field $h=h_c$ the system undergoes a
quantum phase transition at zero temperature: For $h<h_c$ the system is
in an ordered phase which may, or may not, be associated to a
non-vanishing order parameter corresponding to the ground-state expectation of
$S_{\underline{k}}^x$. The order parameter $M_x=\langle
S_{\underline{k}}^x\rangle$ in the case of ferromagnetic order,
and $M_x=(-1)^i\langle S_{\underline{k}}^x\rangle$ in the anti-ferromagnetic case.
In the following, we specialize to the anti-ferromagnetic case; trivial
modifications are needed in the ferromagnetic case.

Single-Qubit Unitary Operations
(SQUOs) are unitary transformations $U_{\underline{k}}$
that leave all spins unaffected but for an arbitrarily chosen
one, say, at site $\underline{k}$, on which the SQUOs act as unitary,
Hermitian, and traceless operators \cite{Giampaolo1}.
One can prove that there exists an element of this class, the
{\it Extremal} SQUO (E-SQUO) $\bar{U}_{\underline{k}}$, such that
the squared Euclidean distance between a
state $\ket{\Psi}$ and its image $\bar{U}_{\underline{k}}
\ket{\Psi}$ under the action of the E-SQUO coincides with the linear
entropy (tangle) $\tau = 2(1-Tr[\rho_k^2])$, where $\rho_k$ denotes
the reduced state of spin $k$ \cite{Coffman,Giampaolo1}.
This entanglement monotone quantifies the entanglement existing in
state $\ket{\Psi}$ between the single-spin block $\underline{k}$
and the remainder of the system. If $\ket{\Psi}$ is the ground state
of a quantum mechanical Hamiltonian $H$, the E-SQUO is
uniquely associated to the aforementioned EXE, defined as
$\Delta E=\langle \bar{U}_{\underline{k}} H \bar{U}_{\underline{k}} \rangle
- \langle H \rangle$. A crucial property of the EXE is that if $H$ is translationally
invariant and $[H, U_{\underline{k}}] \neq 0$ for all SQUOs $U_{\underline{k}}$,
then the ground state is
completely factorized if and only if $\Delta E=0$ \cite{Giampaolo2}.
The generic spin-$1/2$ models \eq{Hamiltonian}, as well as many others,
satisfy this condition. The above theorem can then be applied to identify
the occurrence of factorization points. By definition,
the E-SQUO can be written as $\bar{U}_{\underline{k}}=\bigotimes_{\underline{i}\neq
\underline{k}}\mathbf{1}_{\underline{i}} \otimes O_{\underline{k}}$
where, following Ref.~\cite{Giampaolo1}, $O_{\underline{k}}$ can be
written as
\begin{equation}\label{SQUO}
O_{\underline{k}}^\pm=S_{\underline{k}}^z \cos \theta \pm S_{\underline{k}}^x \sin   \theta \; ,
\end{equation}
where the $\pm$ sign discriminates the two sublattices, corresponding
to the sign of the staggered magnetization $\langle S_{\underline{k}}^x \rangle$.

Let us assume that $H$, \eq{Hamiltonian}, admits a factorized ground state.
Applying the E-SQUO, \eq{SQUO}, to the ground state, exploiting the fact
that for fully factorized states all correlations separate in products of single-site
expectations, and imposing the condition $\Delta E = 0$, one has that
factorization requires the simultaneous occurrence of
\begin{equation}
\label{vanishingcondition}
\tan \theta = \frac{M_x}{M_z} \; ; \; \; \; \; \tan \theta =
\frac{\mathcal{J}_x M_x}{\mathcal{J}_z M_z-h_f} \; ,
\end{equation}
where $h_f$ is the factorizing field, i.e. the value of the external
field for which ground state factorization occurs. The quantities
$\mathcal{J}_\alpha$ are the net interactions reflecting
the type of magnetic order that exists along different axes.
In the antiferromagnetic case,
$\mathcal{J}_x= \sum_{r=1}^\infty (-1)^{r} Z_r J_x^r$ and
$\mathcal{J}_z=\sum_{r=1}^\infty Z_r J_z^r$, where $Z_r$ denotes
the number of sites placed at distance $r$ from a given spin.
Conditions (\ref{vanishingcondition}) and the vanishing of
the tangle $\tau = 1-4(M_x^2+M_z^2)$ yield a closed expression
for the phase $\theta$ as a function of the Hamiltonian parameters and of $h_f$:
\begin{equation}
\label{E-squo}
\cos \theta = \frac{2 h_f}{\mathcal{J}_z-\mathcal{J}_x} \; .
\end{equation}
\eq{E-squo} determines, independently of the magnetizations,
the form of the candidate factorized ground state
$\ket{\Psi_f}$ \cite{Notarella}:
\begin{equation}
\label{groundstate}
\ket{\Psi_f}=\bigotimes_{\underline{i}}\ket{\psi_{2{\underline{i}}}^+}
\ket{\psi_{2{\underline{i}}+1}^-} \; ,
\end{equation}
where $\ket{\psi_{\underline{k}}^\pm}$ are the eigenvectors of $O_{\underline{k}}^\pm$ with
eigenvalue $1/2$.

This far, we have determined the general expression that a factorized ground
state must assume. We are left to establish the conditions for its existence, i.e. the conditions
under which a state of the form \eq{groundstate} is indeed the eigenstate of
$H$, \eq{Hamiltonian}, with the lowest energy. For each pair of spins
${\underline{i}}$ and ${\underline{j}}$ we introduce
the pair Hamiltonian
\begin{equation}
\label{couple}
H_{{\underline{i}}{\underline{j}}}=J_x^r S_{\underline{i}}^x S_{\underline{j}}^x+
J_y^r S_{\underline{i}}^y S_{\underline{j}}^y +J_z^r S_{\underline{i}}^z S_{\underline{j}}^z-
h^r_f (S_{\underline{i}}^z + S_{\underline{j}}^z)\, ,
\end{equation}
where $h^r_f$ is defined by the relation $2 h_f^r = \cos\theta (J_z^r - (-1)^{r} J_x^r)$.
It is immediate to verify that by re-summing the operators
$H_{{\underline{i}}{\underline{j}}}$ over all spin pairs, one reobtains
\eq{Hamiltonian}, with $h=h_f$. Hence, proving that
$\ket{\Psi_f}$ is a simultaneous eigenstate of all pair Hamiltonians
$H_{{\underline{i}}{\underline{j}}}$, implies that it is an eigenstate
of the total Hamiltonian $H$ as well. Inserting the expression
of $\ket{\Psi_f}$ in \eq{couple} yields a set of conditions that must
be satisfied to ensure that $\ket{\Psi_f}$ is an eigenstate of
every pair Hamiltonian $H_{{\underline{i}}{\underline{j}}}$:
\begin{equation}
\label{condition}
- J_y^r+\cos^2\theta J_x^r  + (-1)^r
\sin^2\theta J_z^r=0 \; \; \; \forall r \, ,
\end{equation}
where $\theta$ is given by \eq{E-squo}. By summing over $r$, term
by term, all the relations in \eq{condition}, and solving for
$h_f$, we eventually obtain the exact, general expression of
the factorizing field as a function of the net interactions along
the different axes:
\begin{equation}
\label{factorizing_field}
h_f=\frac{1}{2}\sqrt{\left(\mathcal{J}_x-\mathcal{J}_z\right)
\left(\mathcal{J}_y-\mathcal{J}_z\right)} \; .
\end{equation}
In \eq{factorizing_field} the net interaction along the $y$ axis
depends on the anti-ferromagnetic order on the $x$ axis and, hence,
it is given by 
$\mathcal{J}_y= \sum (-1)^{r} Z_r J_y^r$. We
remark that \eq{factorizing_field} is completely general and holds
for lattices of arbitrary spatial dimension and for interactions
of arbitrary range.

We are left to determine the conditions under which
the factorized eigenstate $\ket{\Psi_f}$ is associated to the
lowest energy eigenvalue.
A general sufficient condition is that every two-site reduced state,
obtained from $\ket{\Psi_f}$ by a partial trace over all sites
except the pair $\{ {\underline{i}}, {\underline{j}}\}$
(which is still a pure state since $\ket{\Psi_f}$ is factorized),
is the ground state of $H_{{\underline{i}}{\underline{j}}}$, for
every pair $\{ {\underline{i}},{\underline{j}}\}$. Given the relation
existing between all the pair Hamiltonians
$H_{{\underline{i}}{\underline{j}}}$ and the total Hamiltonian $H$,
it follows that if a state is associated to the lowest eigenvalue of every $H_{{\underline{i}}{\underline{j}}}$, then it is associated
to the lowest eigenvalue of $H$.
To proceed, we need to distinguish between the various possible cases,
depending on the structure of the couplings $J_\alpha^r$.

\vspace{0.2cm}

\begin{figure}[ht]
\includegraphics[width=7.0cm]{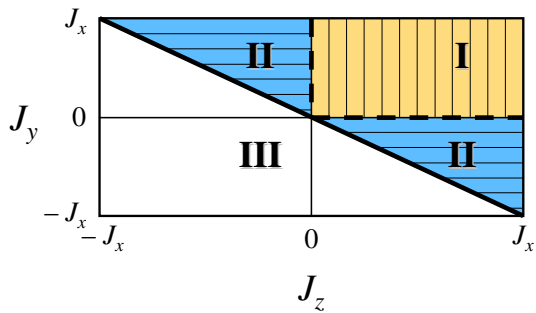}
\caption{ (Color online) Spin-$1/2$ models with short range interactions:
Domains in the space of couplings.
Region I (vertical lines): Domain of parameters for
which ground state factorization was originally identified \cite{Kurmann},
and is established rigorously in the present paper. Region II (horizontal lines):
Domain in which factorization is established for the first time, and rigorously,
in the present paper. Region III (no lines): Domain in which
factorization is rigorously excluded. All plotted quantities are dimensionless.} \label{fig}
\end{figure}

{\noindent \bf Models with short range interactions}.--
By short range, or nearest-neighbor, we mean $J_\alpha^r=0$ for all
$\alpha$ and $r\ge2$. Anti-ferromagnetic order is ensured
by having $J_x^1=1$ and $|J_{y,z}^1|\le 1$. For one-dimensional
models, we have $Z_1=2$, $\mathcal{J}_x=-2$,
$\mathcal{J}_y=-2J_y^1$, and $\mathcal{J}_z=2J_z^1$.
Inserting these quantities in Eqs.~(\ref{factorizing_field},
\ref{E-squo}, \ref{groundstate}) we obtain the expressions for
$\theta$ and for the factorized ground state $\ket{\Psi_f}$. The
explicit expression of the factorizing field is
$h_f^{(d=1)}=\sqrt{(1+J^1_z)(J^1_y+J^1_z)}$ and the energy per site reads $\varepsilon^{(d=1)}=(1/8)(\mathcal{J}_z -\mathcal{J}_y+2)=(1/4)(1+J_z^1+J_y^1)$.
Besides reproducing the original results of Kurmann
{\it et al.} \cite{Kurmann}, the analytic method allows to establish that
ground state factorization occurs for a much larger range of values of the
couplings. Our general framework singles out novel instances of classical-like
ground states already in this simple model, as pictorially sketched in \fig{fig}.

The analytic approach can be extended immediately to
higher-dimensional lattices.
For a square lattice, the number of nearest neighbors is $Z_1=4$, and
hence the net interactions read $\mathcal{J}_{x}=- 4$,
$\mathcal{J}_{y}=-4 J_y^1$, and $\mathcal{J}_{z}=4 J_z^1$.
Therefore, moving from one to two dimensions, we find that
factorization occurs in the same domains of couplings and for
the same value of $\theta$, but at a factorizing field and with
an energy per site that are twice the corresponding quantities in one dimension:
$h_f^{(d=2)}=2h_f^{(d=1)}$ and $\varepsilon^{(d=2)}=2
\varepsilon^{(d=1)}$. These exact results recover, confirm, and extend
recent numerical findings \cite{Roscilde}.

As the resources for numerical simulations scale with the lattice dimension, and
in the absence of analytic approaches, it is not surprising that no study of
factorization in three-dimensional systems was attempted so far. However,
exploiting our novel analytic method, such a study can be carried out exactly and
straightforwardly. Namely, moving from square to cubic lattices we only need to
insert the correct value of the coordination number $Z_1=6$. It is then straightforward
to prove that for three-dimensional models ground state factorization occurs at the
same values of the couplings as in one dimension, but at a factorization
field and with an energy per site that are three times the
corresponding quantities in one dimension:
$h_f^{(d=3)}=3h_f^{(d=1)}$ and $\varepsilon^{(d=3)}=3
\varepsilon^{(d=1)}$.

The ferromagnetic counterparts of the anti-ferromagnetic models
can be immediately recovered performing $\pi/2$-rotation around the
$z$ axis at each lattice site in either one of the two sublattices. Hence,
given a set of couplings $(1,J^1_y,J^1_z)$ for which an anti-ferromagnetic
factorized ground state occurs, there exists a corresponding set of
couplings $(-1,-J^1_y,J^1_z)$ for which a ferromagnetic factorized ground
state occurs at the same value of the factorizing field $h_f$, the same energy
per site, and the same value of $\theta$.


\vspace{0.2cm}

{\noindent \bf Models with finite range interactions}.--
In the case of spin systems with short range interactions,
the set of equations \eq{condition} is automatically verified
by any model that admits a real solution for \eq{E-squo} and \eq{factorizing_field}
\cite{Notarella2}.
This redundancy is removed as soon as one considers interactions
of longer spatial range, because in these cases the net interactions
do not depend on the nearest-neighbor couplings alone.
To illustrate this important point, let us consider ferromagnetic
models in arbitrary spatial dimensions and with non-vanishing interactions
up to a certain finite distance $s$:
$J_{x,y}^{r} < 0 \; \forall r \le s$, $J_{x,y}^{r} = 0 \; \forall r \ge s$,
and $J_z^{r}=0 \; \forall r $.
It is straightforward to verify that a sufficient condition for
ground state factorization is that the ratio of the non vanishing
couplings must satisfy the relation $J_{x}^{r}/J_{x}^{1} = J_{y}^{r}/J_{y}^{1}
\equiv \gamma_{r} \; \forall r \le s$. Otherwise, \eqs{condition} do not admit
solutions.
If this condition is satisfied, we have that the net interactions (that in the
ferromagnetic instance read $\mathcal{J}_{x,y}=\sum_r Z_r J_{x,y}^r$) are
$\mathcal{J}_{x,y} = \Gamma J_{x,y}^{1}$, where $\Gamma = \sum_{r=1}^{s} Z_{r} \gamma_{r}$.
Ground state factorization occurs at $h_{f}^{(s)} = \Gamma (J_x^1 J_y^1)^{1/2}$,
with an energy per site $\varepsilon=[\Gamma(J_x^1+J_y^1)]/8$.
Therefore, at exactly defined ratios of the couplings, systems with finite range
interactions admit fully separable ground states analogous to the ones arising in
the case of systems with only nearest-neighbor couplings.
The factorizing field and the energy per site are increased
exactly by a factor $\Gamma/2$ with respect to the case of models with short
range interactions.

\vspace{0.2cm}

{\noindent \bf Models with infinite range interactions}.--
A very interesting limiting case is given by models with infinite range interactions,
such as the fully connected or Lipkin-Meshkov-Glick (LMG) model
\cite{Lipkin}, which is obtained in the limit of diverging $s$ and couplings
of the form $J_{x,y}^r = J_{x,y} = 2 \Delta_{x,y}/N \; \forall r$. The scaling with the number
of lattice sites $N$ ensures that the net interactions converge to a finite
value in the thermodynamic limit: $\mathcal{J}_{x,y} \rightarrow 2 \Delta_{x,y}$.
It is then rather straightforward to solve \eqs{condition} exactly and prove rigorously that
the ground state of the LMG model is a fully factorized ferromagnetic state for
$\Delta_x = 1$; $0 \le \Delta_y \le 1$; $h_f = \sqrt{\Delta_{y}}$ at
$\theta = \arccos h_{f}$; and an energy per site
$\varepsilon = (1 + \Delta_{y})/4$. These results justify
rigorously recent numerical findings \cite{Dusuel}.

\vspace{0.2cm}

{\noindent \bf Comments}.-- We have introduced
a simple and general analytic approach to the exact determination
of factorized ground states in quantum spin systems. We have applied
the scheme to spin-$1/2$ models with general anisotropic Heisenberg-like
interactions of arbitrary range and for lattices of arbitrary
dimensions. Besides the rigorous derivation of the few known,
mainly numerical, results, we have showed that our method
allows to determine exactly novel classes of factorization points
in various models, generally non-exactly solvable, for different
lattice dimensions and for different interaction ranges. These 
novel exact solutions of non exactly solvable models are obtained 
for nontrivial sets of values of the Hamiltonian parameters. 
Furthermore, according to the general theorem by Kurmann {\em et al.}
on factorization \cite{Kurmann},
given any Hamiltonian of the form \eq{Hamiltonian} with generic
spin $ S > 1/2$, the ground state of the system is factorized
at the same value of the external field $h=h_f$ [\eq{factorizing_field}],
at which factorization occurs in the corresponding spin-$1/2$ model.
Therefore, the method and the results derived in the present
paper are straightforwardly generalized to interacting systems with
arbitrary value of the spin which are endowed with the same Hamiltonian
structures as in the spin-$1/2$ case. Further applications to
other systems, defined, e.g., on ladders and coupled planes,
or to models with frustration and in complex geometries, can be in principle
carried out by suitably adapting and specializing the general framework
introduced in the present work. From a conceptual standpoint, the method
realizes a rigorous and analytic implementation of concepts motivated by
quantum information theory to obtain genuinely new insights on founding
open questions of condensed matter physics.

\vspace{0.2cm}

{\noindent \bf Acknowledgements}.-- We thank
T. Roscilde and P. Verrucchi for useful discussions.
We acknowledge financial support from MIUR under PRIN
National Project 2005 and from CNR-INFM Coherentia.


\begin{thebibliography}{99}
\bibitem{Notarella}
If one does not require the vanishing of the
EXE, $\theta$ and the ground state \eq{groundstate}
are expressed as a simple function of
the magnetizations $M_\alpha$'s alone. Imposing the conditions
under which \eq{groundstate} is an eigenstate of the Hamiltonian,
one merely obtains a relation between $h_f$ and the $M_\alpha$'s.
If the analytic expression of at least one of the $M_\alpha$'s is
known, it is possible to obtain an expression of the factorization
point $h_f$ as a function of the Hamiltonian parameters. But this
occurs only in few special cases.
\bibitem{Notarella2}
Notice that while \eq{condition} involves only the couplings at a fixed
distance $r$, \eq{E-squo} and \eq{factorizing_field} involve the net
interactions. Hence, in models with short range interactions
$(J_\alpha^r=0 \; \forall \alpha \;, r \ge 2)$, if \eq{E-squo} and
\eq{factorizing_field} have two real solutions, then \eqs{condition} is
satisfied $\forall r$. On the contrary, for models with finite and long
range interactions there may be cases when \eq{E-squo} and
\eq{factorizing_field} have two real solutions but \eqs{condition}
are not verified $\forall r$ and thus factorization cannot occur.
\end{thebibliography}
\end{document}